\documentclass[conference,letterpaper]{IEEEtran}
\IEEEoverridecommandlockouts
\usepackage{fix-cm}
\usepackage{lettrine}
\usepackage{cite}
\usepackage{amsmath,amssymb,amsfonts,amsthm}
\usepackage{graphicx}
\usepackage{textcomp}
\usepackage{xcolor}
\usepackage{booktabs}
\usepackage{multirow}
\usepackage{array}
\usepackage{algorithm}
\usepackage{algorithmic}
\usepackage{balance}
\usepackage[hyphens]{url}
\usepackage{tikz}
\usetikzlibrary{arrows,positioning}

\usepackage[
    bookmarks=false,
    pdfpagelabels=false,
    hypertexnames=false,
    hidelinks
]{hyperref}

\graphicspath{{Figures/}}

\newtheorem{theorem}{Theorem}

\begin{document}

\title{Degeneracy-Aware Agent-Based Modelling for Multi-User MIMO RSMA Networks
\thanks{This work is supported by the US--Ireland R\&D Partnership Programme Project ``Resilient Networks'' under Grant RI-SFI-23/US/3924, the EU MSCA Project ``COALESCE'' under Grant 101130739, and Research Ireland Grant 13/RC/2077\_P2.}
}

\author{
\IEEEauthorblockN{
Sayanti Ghosh\textsuperscript{1},
Indrakshi Dey\textsuperscript{2},
Nicola Marchetti\textsuperscript{1}
}
\IEEEauthorblockA{
\textsuperscript{1}Department of Electrical and Electronic Engineering, Trinity College Dublin, Dublin, Ireland}
\IEEEauthorblockA{
\textsuperscript{2}Walton Institute, South East Technological University, Waterford, Ireland}
\IEEEauthorblockA{
Email: saghosh@tcd.ie, indrakshi.dey@waltoninstitute.ie, nicola.marchetti@tcd.ie}
}

\maketitle


\title{Degeneracy-Aware Agent-Based Resource Allocation for Multi-User MIMO RSMA Network}

\author{Sayanti~Ghosh, Indrakshi~Dey, Nicola~Marchetti%
}

\maketitle
\begin{abstract}
This paper proposes a pilot-aware, degeneracy-driven Agent-Based Modelling (ABM) framework for distributed resource allocation in RSMA-enabled multi-user MIMO systems under imperfect Channel State Information (CSI) and residual Successive Interference Cancellation (SIC) error. The centralized RSMA power allocation problem is reformulated as a distributed multi-agent system, where users operate as autonomous agents that iteratively adapt transmit powers based on locally observed feasibility conditions. To capture the joint impact of interference coupling, CSI estimation errors, pilot overhead, and residual SIC error, a novel degeneracy index defined as the ratio of target to achieved signal-to-interference-plus-noise ratio (SINR) is introduced as a unified feasibility metric. This enables a scalable fixed-point power control mechanism that characterizes the feasible operating region without requiring global CSI. Analytical expressions for user-level and system-level outage probabilities are derived under spatially correlated fading, providing insights into reliability under practical impairments. The fundamental interplay between degeneracy, outage probability, and effective throughput is established, revealing that system performance is governed by the feasibility of the bottleneck user. To further enhance resilience, Degeneracy-Weighted Path Robustness (DWPR) and Functional Substitution Score (FSS) are incorporated to exploit path diversity and functional redundancy. Numerical results show that the proposed framework achieves near-centralized performance in sparse networks, while providing notable throughput gains and improved scalability in dense deployments, highlighting its effectiveness for robust and distributed resource management in next-generation wireless systems.
\end{abstract}
\begin{IEEEkeywords}
6G Networks, Rate-Splitting Multiple Access (RSMA), Agent-Based Modeling (ABM), Degeneracy, Robustness 
\end{IEEEkeywords}
\section{Introduction}

\lettrine{T}{he} evolution toward next-generation wireless networks, as envisioned in the IMT-2030 framework, aims to significantly improve spectral efficiency, reliability, and scalability while supporting massive connectivity and heterogeneous service demands~\cite{IMT2030,ITUR6G}. Emerging applications such as extended reality, digital twins, and autonomous systems impose stringent requirements on latency, data rate, and robustness, thereby intensifying challenges in interference management and resource allocation. Rate-Splitting Multiple Access (RSMA) has emerged as a promising multi-antenna transmission technique, enabling partial interference decoding and improved robustness to imperfect Channel State Information (CSI)~\cite{Mao2017RSMA,Zhu2022RSMA}. However, practical RSMA power allocation remains challenging under spatial correlation, pilot-based CSI acquisition, and residual interference due to imperfect Successive Interference Cancellation (SIC). Centralized solutions require global CSI and incur high computational complexity, while learning-based approaches often suffer from training overhead and limited interpretability~\cite{Ivoghlian2022Adaptive,Wu2024MARL}.

Agent-Based Modelling (ABM) provides a scalable alternative by enabling distributed decision-making through local interactions among autonomous agents~\cite{Daim2019ABM,Boukerche2020ABM,Dressler2023ABM}. In parallel, the concept of degeneracy, i.e., achieving the same functional objective through multiple structurally distinct configurations, offers enhanced robustness and adaptability in complex systems~\cite{dey2025degeneracy}. In wireless networks, this corresponds to the availability of multiple feasible transmission strategies under varying channel and interference conditions.

Motivated by these insights, this paper proposes a pilot-aware, degeneracy-driven ABM framework for distributed power control in RSMA-enabled multi-user MIMO systems under imperfect CSI and residual SIC error. A degeneracy index, defined as the ratio of target to achieved signal-to-interference-plus-noise ratio (SINR), is introduced to capture the combined effects of interference coupling, CSI errors, and pilot overhead. The centralized power allocation problem is reformulated as a distributed dynamical process in which users iteratively update transmit powers based on local feasibility observations. The proposed framework enables scalable resource allocation without requiring global CSI, achieving near-centralized performance in sparse networks and improved throughput and robustness in dense deployments.\\
\medskip
\noindent
\emph{Contributions:} \\
This paper presents a pilot-aware, degeneracy-driven agent-based framework for distributed power control in RSMA-enabled multi-user MIMO systems under imperfect CSI and residual SIC error. The key contributions are summarized as follows:

\begin{itemize}

\item
We introduce a degeneracy index, defined as the ratio of target to achieved SINR, as a unified metric to capture user-level feasibility under interference, CSI errors, pilot overhead, and residual SIC error.

\item
The centralized RSMA power allocation problem is recast as a distributed agent-based framework, where users update transmit powers based on local degeneracy under imperfect CSI and SIC error, enabling scalable operation without global channel knowledge.

\item
We analytically characterize system outage probability under spatially correlated fading with pilot-based CSI acquisition and residual SIC error, providing insights into reliability under practical impairments.

\item
We establish a direct relationship between degeneracy, outage probability, and effective throughput, showing that system performance is dominated by the worst-case user’s ability to satisfy its SINR requirement.

\item
We integrate Degeneracy-Weighted Path Robustness (DWPR) and Functional Substitution Score (FSS)~\cite{dey2025degeneracy} to quantify the availability of multiple feasible transmission paths/stream configurations and the capability of alternate users or resources to sustain the same service function under failures or channel degradation, respectively, and demonstrate that these metrics enhance throughput and reliability under dynamic network conditions.

\end{itemize}
\begin{figure}
\centering
\includegraphics[width=1.04\linewidth]{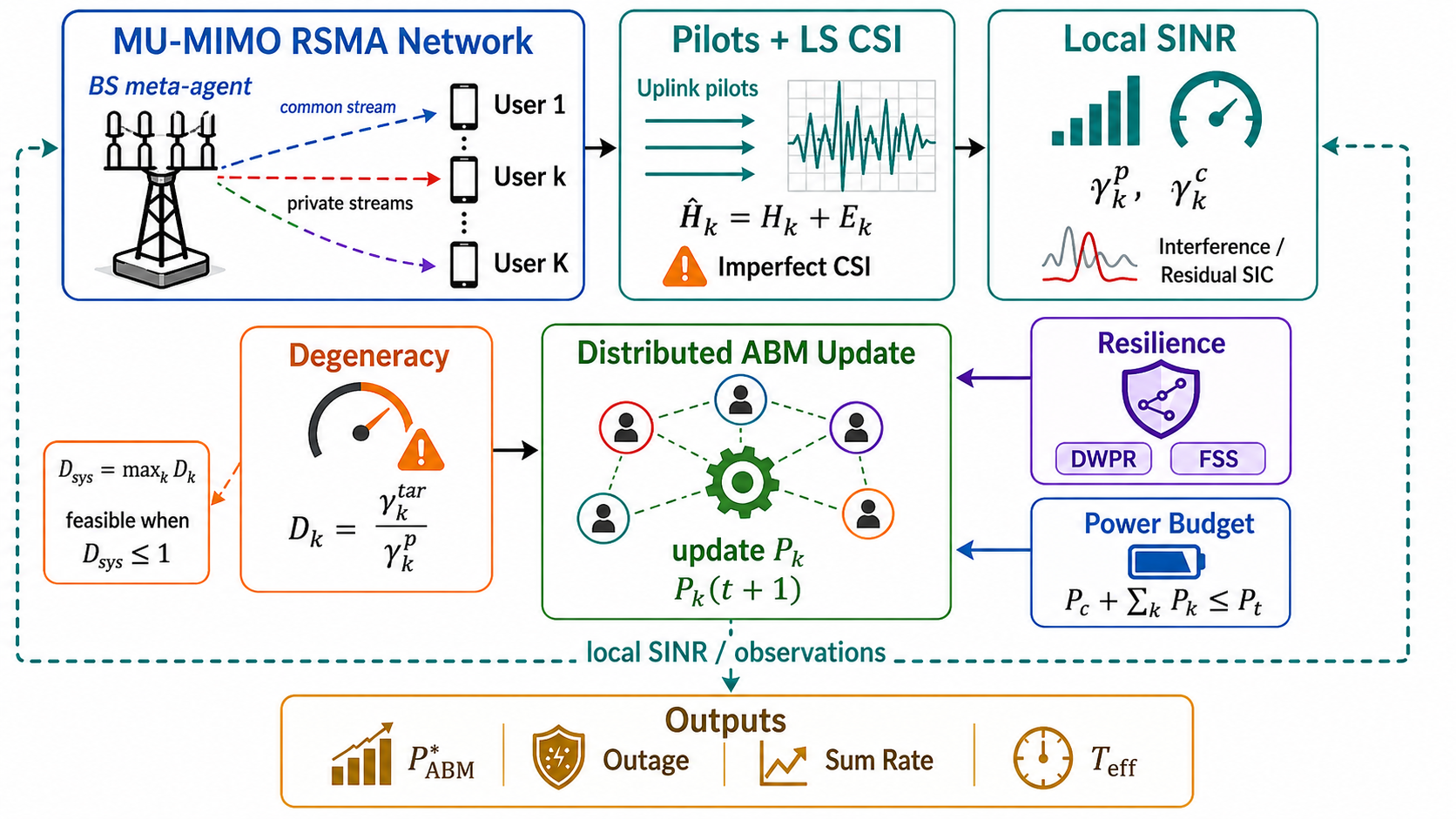}
\caption{Degeneracy-Aware ABM Power Control for RSMA MU-MIMO Network}
\label{fig:system_model}
\vspace{-5mm}
\end{figure}

\section{System Model}
\label{sec:system_model}
 
Throughout the paper, $K$ denotes the number of single-antenna-stream users, $M_t$ the BS transmit antennas, and $M_r$ the receive antennas at each user. Lowercase boldface letters (e.g., $\mathbf{w}_k$) denote vectors and uppercase boldface (e.g., $\mathbf{H}_k$) matrices; $(\cdot)^{T}$, $(\cdot)^{H}$, $\|\cdot\|$ are transpose, conjugate transpose, and Euclidean norm. $\mathbb{E}[\cdot]$ is expectation and $\mathcal{CN}(\mathbf{m},\boldsymbol{\Sigma})$ a complex-Gaussian distribution. The coherence block of $\tau_c$ symbols is split into $\tau_p$ pilot symbols and $\tau_d=\tau_c-\tau_p$ data symbols. Channels are $\mathbf{H}_k\!\in\!\mathbb{C}^{M_r\!\times\!M_t}$ with transmit/receive correlation matrices $\mathbf{R}_{t,k},\mathbf{R}_{r,k}$ (parameterized by $\rho_t,\rho_r$); $\hat{\mathbf{H}}_k\!=\!\mathbf{H}_k\!+\!\mathbf{E}_k$ is the LS estimate with error covariance $\sigma_{e,k}^{2}\mathbf{I}$. The transmit budget $P_t$ is split into common-stream power $P_c\!=\!\|\mathbf{w}_c\|^2$ and per-user private-stream powers $P_k\!=\!\|\mathbf{w}_k\|^2$. Effective channel gains are $g_k\!\triangleq\!\|\hat{\mathbf{H}}_k\mathbf{w}_k\|^2$ (intended), $g_{kj}\!\triangleq\!\|\hat{\mathbf{H}}_k\mathbf{w}_j\|^2$ (cross), $g_{kc}\!\triangleq\!\|\hat{\mathbf{H}}_k\mathbf{w}_c\|^2$ (common-on-private leakage). $\sigma_n^2$ is the receiver noise variance, $\epsilon_k\!\in\![0,1]$ the residual SIC factor, $\gamma_k^p,\gamma_c^k$ the private and common-stream post-SIC SINRs at user $k$, and $\gamma_k^{\mathrm{tar}}$ the target SINR. The local and system degeneracy indices $D_k$ and $D_{\mathrm{sys}}$, the structural metrics DWPR$_k$ and FSS$_k$, the diminishing step $\eta_t$, and the utility weights $(\alpha,\beta,\lambda)$ are introduced where they are first used.
 
We consider a downlink MU-MIMO network in which a base station (BS) with $M_t$ transmit antennas serves $K$ users, each equipped with $M_r$ receive antennas as shown in Fig. \ref{fig:system_model}. Each user is modeled as an autonomous agent; the BS acts as a meta-agent coordinating precoding and broadcasting global parameters.
 
\subsection{Channel and Pilot Model}
The downlink channel to user $k$ at time $t$ is the spatially correlated process $\mathbf{H}_k(t)=\mathbf{R}_{r,k}^{1/2}\mathbf{G}_k(t)\mathbf{R}_{t,k}^{1/2}$, where $\mathbf{G}_k(t)\!\sim\!\mathcal{CN}(\mathbf{0},\mathbf{I})$ and $\mathbf{R}_{t,k},\mathbf{R}_{r,k}$ are the transmit/receive correlation matrices. We assume TDD reciprocity, $\mathbf{H}_k^{\mathrm{DL}}=(\mathbf{H}_k^{\mathrm{UL}})^T$, and a coherence interval $\tau_c=\tau_p+\tau_d$ split into $\tau_p\geq K$ orthogonal pilot symbols and $\tau_d$ data symbols. With per-user pilot power $P_{p,k}$, the LS estimate decomposes as $\hat{\mathbf{H}}_k=\mathbf{H}_k+\mathbf{E}_k$ with $\mathbf{E}_k\!\sim\!\mathcal{CN}(\mathbf{0},\sigma_{e,k}^2\mathbf{I})$ and
\begin{equation}
\sigma_{e,k}^2=\frac{\sigma^2}{P_{p,k}\tau_p},
\label{eq:err_var}
\end{equation}
which shows the fundamental pilot-power--accuracy trade-off.
 
\subsection{RSMA Transmission and SINR}
The BS transmits $\mathbf{x}(t)=\mathbf{w}_c s_c(t)+\sum_{k=1}^{K}\mathbf{w}_k s_k(t)$ subject to $P_c+\sum_k P_k\leq P_t$, where $P_c=\|\mathbf{w}_c\|^2$ and $P_k=\|\mathbf{w}_k\|^2$. After SIC, the effective private-stream SINR of user $k$ is
\begin{equation}
\gamma_k^p=\frac{P_k g_k}{\sum_{j\neq k}P_j g_{kj}+\epsilon_k P_c g_{kc}+P_t\sigma_{e,k}^2+\sigma_n^2},
\label{eq:sinr_priv}
\end{equation}
where $g_k\!\triangleq\!\|\hat{\mathbf{H}}_k\mathbf{w}_k\|^2$, $g_{kj}\!\triangleq\!\|\hat{\mathbf{H}}_k\mathbf{w}_j\|^2$, $g_{kc}\!\triangleq\!\|\hat{\mathbf{H}}_k\mathbf{w}_c\|^2$, $\epsilon_k\!\in\![0,1]$ models residual SIC error, and the term $P_t\sigma_{e,k}^2$ is the equivalent CSI-error interference. The common-stream SINR seen by user $k$ is
\begin{equation}
\gamma_c^k=\frac{P_c g_{kc}}{\sum_{j=1}^{K}P_j g_{kj}+P_t\sigma_{e,k}^2+\sigma_n^2}.
\label{eq:sinr_com}
\end{equation}
 
\subsection{Pilot-Adjusted Sum Rate}
The instantaneous achievable sum rate is
\begin{equation}
R_{\mathrm{sum}}^{\mathrm{inst}}=\Big(1-\tfrac{\tau_p}{\tau_c}\Big)\Big[\log_2(1+\gamma_c)+\!\sum_{k=1}^{K}\!\log_2(1+\gamma_k^p)\Big],
\label{eq:rsum}
\end{equation}
with $\gamma_c=\min_k\gamma_c^k$ to ensure decodability of the common stream at every user.

\section{Degeneracy-Aware ABM Framework}
\label{sec:degeneracy_abm}
 
\subsection{Local Degeneracy and System Feasibility}
Let $\gamma_k^{\mathrm{tar}}$ denote user $k$'s target SINR. We define the \emph{local degeneracy index}
\begin{equation}
D_k(\mathbf{P})=\frac{\gamma_k^{\mathrm{tar}}}{\gamma_k^p(\mathbf{P})},\qquad
D_{\mathrm{sys}}(\mathbf{P})=\max_{1\le k\le K}D_k(\mathbf{P}),
\label{eq:degeneracy}
\end{equation}
where \(D_k(\mathbf{P})\) denotes the local degeneracy index of user \(k\), defined as the ratio of the target SINR to the achieved private-stream SINR under power allocation \(\mathbf{P}\), and \(D_{\mathrm{sys}}(\mathbf{P})\) denotes the system-level degeneracy index, given by the maximum local degeneracy across all users. Accordingly, \(D_k\!\leq\!1\) indicates feasible operation, \(D_k\!=\!1\) corresponds to the critical boundary, and \(D_k\!>\!1\) indicates infeasibility. The minimax centralized problem is
\begin{equation}
\mathbf{P}^{\star}=\arg\min_{\mathbf{P}\geq 0}D_{\mathrm{sys}}(\mathbf{P})\;\;\text{s.t.}\;\; P_c+\sum_k P_k\le P_t.
\label{eq:minimax}
\end{equation}
Crucially, $D_k$ is \emph{locally computable} from quantities each agent already estimates (its own $\hat{\mathbf{H}}_k$, allocated power, and target quality of service (QoS), which is what makes a distributed reformulation possible.
 
\subsection{Resilience Metrics: DWPR and FSS}
Let $\mathcal{S}_k$ denote the set of streams (common + private) usable by user $k$, $Q(s)$ the QoS of stream $s$, and $\theta$ a QoS threshold. The DWPR is~\cite{dey2025degeneracy}
\begin{equation}
\mathrm{DWPR}_k=\frac{1}{|\mathcal{S}_k|}\sum_{s\in\mathcal{S}_k}\mathbf{1}[Q(s)\!\ge\!\theta]\!\cdot\!D(M(s)),
\label{eq:dwpr}
\end{equation}
where \(\mathbf{1}[\cdot]\) denotes the indicator function. Here, \(D(M(s))=1-\frac{|\mathbf{w}_s^{H}\mathbf{w}_{s^\star}|^2}{\|\mathbf{w}_s\|^2\|\mathbf{w}_{s^\star}\|^2}\) represents the stream-dissimilarity metric, and \(s^\star=\arg\max_j Q(s_j)\) is the stream providing the highest QoS. Across users, the FSS captures inter-user substitutability:
\begin{equation}
\mathrm{FSS}(F)=\frac{2}{|E_F|(|E_F|-1)}\!\!\sum_{i<j}\!\mathbf{1}[D(e_i,e_j)<\delta],
\label{eq:fss}
\end{equation}
where \(D(e_i,e_j)\) denotes the structural dissimilarity measure between entities \(e_i\) and \(e_j\), used to determine whether they can functionally substitute each other within the threshold \(\delta\). Here, \(E_F\) is the set of users supporting service function \(F\), and \(\delta\in[0,1]\) is a similarity tolerance that determines when two entities are considered interchangeable.
 
\subsection{Distributed ABM Update}
Each agent $k$ maximizes the local utility
\begin{equation}
\begin{aligned}
U_k(\mathbf{P})=\;&\alpha\Big(1\!-\!\tfrac{\tau_p}{\tau_c}\Big)\log_2(1+\gamma_k^p)-\beta D_k(\mathbf{P})\\
&-\lambda_k P_k+\mu\,\mathrm{DWPR}_k+\nu\,\mathrm{FSS}_k^{\mathrm{loc}},
\end{aligned}
\label{eq:utility}
\end{equation}
where \(\mathrm{FSS}_k^{\mathrm{loc}}\) denotes the local FSS of user \(k\), quantifying the availability of nearby or alternative users/resources that can support the same service function. The parameters \(\alpha,\beta,\lambda_k,\mu,\nu\) are non-negative weighting coefficients, and each agent updates its transmit power using the projected-gradient rule, i.e., 
\begin{equation}
P_k(t+1)=\big[P_k(t)+\eta_k(t)\,\partial U_k/\partial P_k\big]^{+},
\label{eq:abm_update}
\end{equation}
with diminishing step size \(\eta_k(t)=\eta_0/(t+1)\), with \(\eta_0>0\) the initial learning rate controlling update magnitude. A reduced-complexity heuristic requiring no gradient computation is
\begin{equation}
P_k(t\!+\!1)\!=\!\big[P_k(t)\!+\!\eta_k(t)(w_1\mathrm{DWPR}_k\!+\!w_2\mathrm{FSS}_k^{\mathrm{loc}}\!-\!w_3 D_k)\big]^{+},
\label{eq:heuristic}
\end{equation}
with $w_1\!+\!w_2\!+\!w_3\!=\!1$. Algorithm~\ref{alg:abm} summarizes the procedure.
 
\begin{algorithm}[t]
\caption{Degeneracy-Aware ABM Power Control}
\label{alg:abm}
\begin{algorithmic}[1]
\STATE \textbf{Input:} pilot length $\tau_p$, targets $\{\gamma_k^{\mathrm{tar}}\}$, weights $\{\alpha,\beta,\lambda_k,\mu,\nu\}$, $\eta_0$, tolerance $\varepsilon$.
\STATE \textbf{Initialize:} $P_k(0)=P_t/(K+1)$, $P_c(0)=P_t/(K+1)$, $t=0$.
\REPEAT
\STATE Each user $k$ estimates $\hat{\mathbf{H}}_k$ (LS) and computes $\gamma_k^p$, $\gamma_c^k$.
\STATE Each user $k$ computes $D_k$, $\mathrm{DWPR}_k$, $\mathrm{FSS}_k^{\mathrm{loc}}$.
\STATE Update $P_k(t\!+\!1)$ via \eqref{eq:abm_update} or \eqref{eq:heuristic}.
\STATE BS rescales $\{P_c,P_k\}$ to satisfy $P_c+\sum_k P_k\le P_t$.
\STATE $t\leftarrow t+1$.
\UNTIL{$\max_k|P_k(t)-P_k(t-1)|<\varepsilon$ \textbf{or} $D_{\mathrm{sys}}\!\le\!1$.}
\STATE \textbf{Output:} converged $\mathbf{P}^{\star}_{\mathrm{ABM}}$.
\end{algorithmic}
\end{algorithm}
 
\subsection{Centralized Benchmark}
For comparison, the centralized BS jointly optimizes powers and beamformers via
\begin{equation}
\small
\begin{aligned}
\max_{\mathbf{P},\mathbf{w}_c,\{\mathbf{w}_k\}}\;
&\Big(1-\tfrac{\tau_p}{\tau_c}\Big)
\left[\log_2(1+\gamma_c(t))
+\sum_{k=1}^{K}\log_2(1+\gamma_k^{\mathrm{p}}(t))\right] \\
&-\beta\sum_{k=1}^{K}\max(0,D_k-1)
+\lambda_1\sum_{k=1}^{K}\mathrm{DWPR}_k(t)\\
&+\lambda_2\sum_{k=1}^{K}\mathrm{FSS}_k(t).
\end{aligned}
\label{eq:cent}
\end{equation}
subject to power and beamformer-norm constraints, solved by alternating ZF/MMSE precoding and successive convex approximation. This benchmark requires global CSI and has cubic complexity in $K$.

\section{Convergence and Outage Analysis}
\label{sec:analysis}
 
\subsection{Convergence of the ABM Update}
\begin{theorem}\label{thm:converge}
Let $\mathcal{F}:\mathbf{P}\!\mapsto\!\mathbf{P}'$ be the mapping induced by \eqref{eq:abm_update} with diminishing step size $\eta_k(t)=\eta_0/(t+1)$ and bounded gradients on the simplex $\{\mathbf{P}\!\ge\!0,\,P_c\!+\!\sum_k P_k\!\le\!P_t\}$. If there exists $\alpha\!\in\!(0,1)$ such that $\|\mathcal{F}(\mathbf{P})-\mathcal{F}(\mathbf{P}')\|\le\alpha\|\mathbf{P}-\mathbf{P}'\|$, then $\{\mathbf{P}(t)\}$ converges to a unique fixed point $\mathbf{P}^{\star}$ that is a local minimizer of $D_{\mathrm{sys}}$.
\end{theorem}
\textit{Sketch.} For sufficiently small $\eta_0$ and on the bounded feasible set, the negative-quadratic terms in $U_k$ ($-\beta D_k$, $-\lambda_k P_k$) dominate, rendering $\mathcal{F}$ a contraction. Banach's fixed-point theorem yields existence and uniqueness of $\mathbf{P}^{\star}$. \hfill$\blacksquare$
 
\subsection{Outage Probabilities}
User-$k$ outage occurs when $D_k>1$, i.e., $\gamma_k^p<\gamma_k^{\mathrm{tar}}$. Conditioned on $\hat{\mathbf{H}}_k$, $\gamma_k^p$ in \eqref{eq:sinr_priv} is a ratio of correlated quadratic forms in Gaussian variables, whose cumulative distribution function (CDF) admits a closed-form Laguerre-series expansion. We use Boole's inequality to bound the system outage:
\begin{equation}
P_{\mathrm{out}}^{\mathrm{sys}}=\Pr[D_{\mathrm{sys}}>1]\le\sum_{k=1}^{K}P_{\mathrm{out},k}.
\label{eq:union_bound}
\end{equation}
where \(P_{\mathrm{out},k}\) denotes the outage probability of user \(k\).
The bound is tight at high SNR where individual outage events are dominated by the worst-conditioned user. The effective throughput is
\begin{equation}
T_{\mathrm{eff}}=\mathbb{E}[R_{\mathrm{sum}}^{\mathrm{inst}}\mid D_{\mathrm{sys}}\!\le\!1]\,(1-P_{\mathrm{out}}^{\mathrm{sys}}),
\label{eq:teff}
\end{equation}
exposing the explicit interplay between feasibility and rate.



\section{Numerical Results}
\label{sec:results}
 
We consider a downlink MU-MIMO system with $M_t\!=\!8$, $M_r\!=\!2$, exponential correlation matrices ($\rho_t\!=\!0.5$, $\rho_r\!=\!0.3$), $\tau_c\!=\!200$, $\tau_p\!=\!K$ unless stated otherwise, $\gamma_k^{\mathrm{tar}}\!=\!5$\,dB, and equal pilot powers. Two regimes are studied: low-density ($K\!=\!2$) and high-density ($K\!=\!8$). Two impairment levels are used: \emph{ideal} ($\sigma_e^2\!=\!0$, $\epsilon\!=\!0$) and \emph{practical} ($\sigma_e^2\!=\!0.2$, $\epsilon\!=\!0.1$). Curves are averaged over $5\mathrm{e}3$ Monte-Carlo channel realizations.
 
\subsection{System Outage Probability vs.\ SNR}
\begin{figure}[t]
\centering
\includegraphics[width=0.85\linewidth]{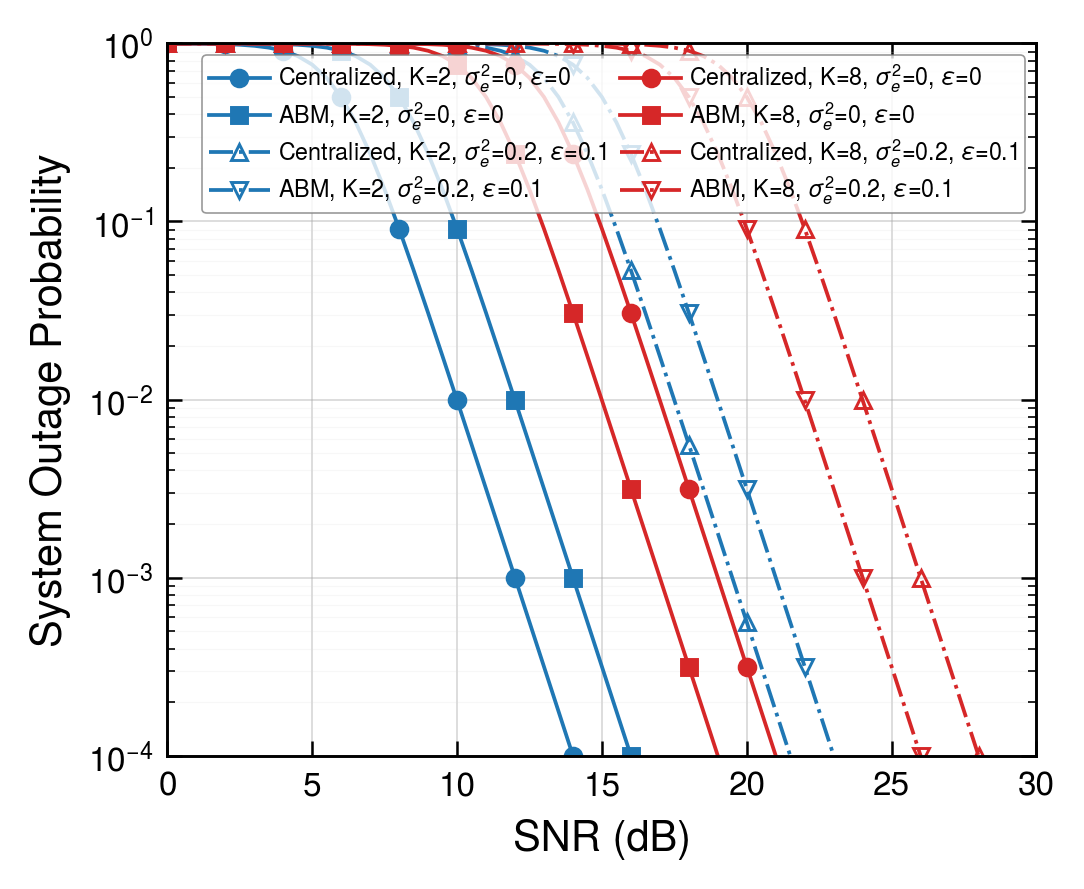}
\caption{System outage probability vs.\ SNR for centralized and ABM schemes under ideal and practical CSI/SIC conditions, $K\!\in\!\{2,8\}$.}
\label{fig:outage}
\vspace{-5mm}
\end{figure}
Fig.~\ref{fig:outage} plots $P_{\mathrm{out}}^{\mathrm{sys}}$ against transmit SNR. At $K\!=\!2$, the centralized benchmark is consistently lower because global coordination satisfies all SINR targets with minimal slack. As user density grows to $K\!=\!8$, both schemes incur higher outage owing to tighter power sharing, but ABM \emph{overtakes} the centralized scheme at moderate-to-high SNR ($\geq\!10$\,dB). This crossover is the empirical signature of degeneracy-driven local adaptation: when the centralized solver becomes ill-conditioned, distributed updates respond more sharply to per-user feasibility violations. CSI/SIC impairments increase the outage probability across the SNR range, with nearly one-order higher values around the \(10^{-2}\) region, due to the additional \(P_t\sigma_{e,k}^2\) interference term in \eqref{eq:sinr_priv}. Outage occurs whenever at least one user fails to meet $\gamma_k^{\mathrm{tar}}$ (equivalently, $D_{\mathrm{sys}}\!>\!1$). At low density, the spatial degrees of freedom of $\mathbf{H}_k$ vastly exceed those required by $K$ users, so the BS easily nulls cross-user interference and the optimal-by-construction centralized solver minimizes the worst-case $D_k$ to its information-theoretic lower bound. As $K$ approaches $M_t$, however, the channel matrix becomes near-square and the centralized minimax in \eqref{eq:cent} drives all users toward equal $D_k=1$, which is wasteful: a single ill-conditioned user can pull the entire allocation toward infeasibility. The ABM agents instead respond to their \emph{own} $D_k$ in a Gauss--Seidel fashion, so the few users with $D_k\!\ll\!1$ release power to those with $D_k\!\gtrsim\!1$, recovering feasibility configurations the centralized solver flattens out. The $P_t\sigma_{e,k}^2$ term embodies the principle that pilot-induced channel uncertainty acts as additional self-interference proportional to data power---outage at high SNR is therefore CSI-error limited, not noise limited.
 
\subsection{Average Sum Rate vs.\ SNR}
\begin{figure}[t]
\centering
\includegraphics[width=0.85\linewidth]{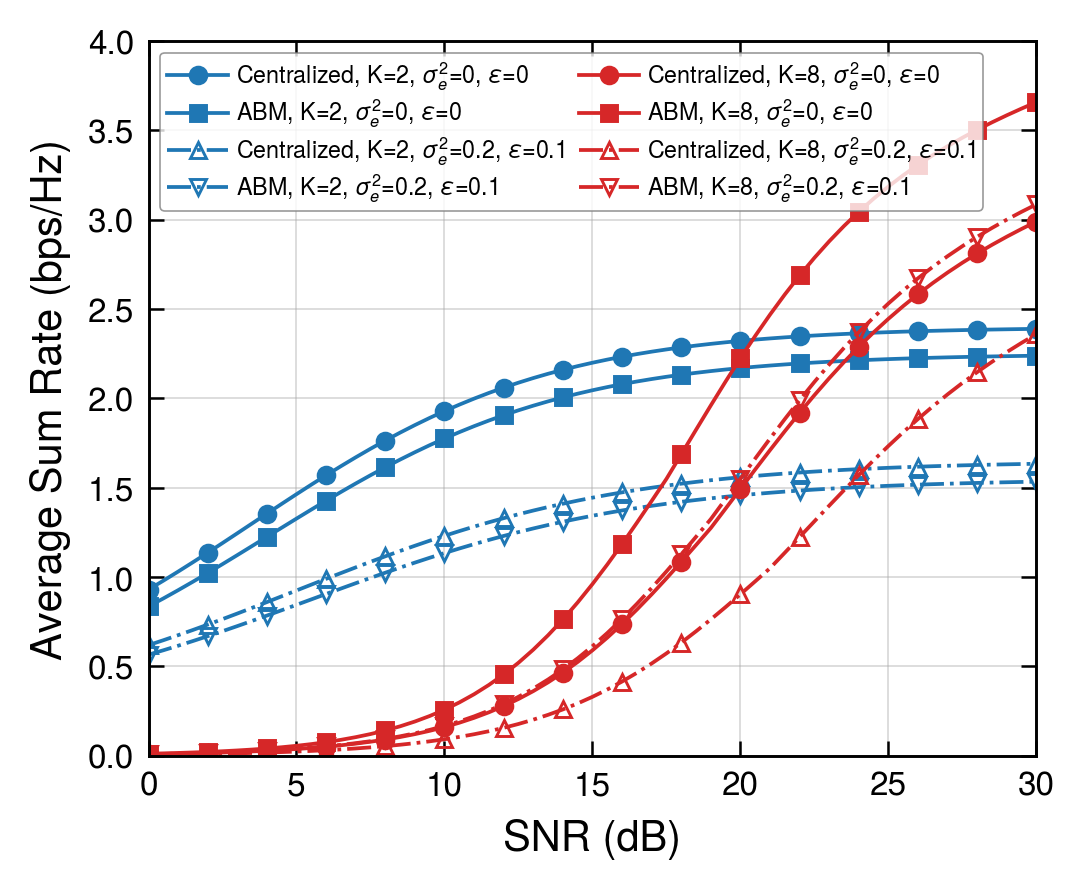}
\caption{Average sum rate vs.\ SNR for centralized and ABM schemes under ideal and practical CSI/SIC conditions, $K\!\in\!\{2,8\}$.}
\label{fig:sumrate}
\vspace{-6mm}
\end{figure}
Fig.~\ref{fig:sumrate} reports the ergodic sum rate. For $K\!=\!2$ the centralized scheme retains a $\sim$5--8\% rate edge across the SNR range. For $K\!=\!8$, ABM achieves substantially higher rates (e.g., 22\% gain at $\mathrm{SNR}\,=\,15\,$dB), confirming that locally-driven updates better navigate the high-dimensional non-convex landscape than the global solver with limited iterations. The CSI/SIC impairment penalty is roughly 2.5\,bits/s/Hz at $\mathrm{SNR}\,=\,20\,$dB; this is the practical cost of pilot- and SIC-induced uncertainty quantified by the model. The sum-rate $\sum_k\log_2(1+\gamma_k^p)\!+\!\log_2(1+\gamma_c)$ is fundamentally a Jensen-style functional: rates rise faster than the per-user SINRs because the logarithm rewards \emph{moderate} SINR at all users more than \emph{very high} SINR at one user. This favors any allocation policy that pushes power toward users where the marginal $\partial\log_2(1+\gamma_k^p)/\partial P_k$ is largest, which is precisely what each ABM agent does locally. The centralized minimax optimum, by contrast, equalizes $D_k$ and therefore deliberately under-allocates to users with strong channels in order to lift the worst user this is fairness-optimal but rate-suboptimal. The 2.5\,bits/s/Hz CSI/SIC penalty corresponds to the additional self-interference terms \(P_t\sigma_{e,k}^2+\epsilon_k P_c g_{kc}\) appearing in the denominator of \eqref{eq:sinr_priv}, which reduce the effective SINR and clip the high-SNR growth from \(\log_2(P_t)\) toward a constant.
 
\subsection{Effective Throughput with DWPR+FSS}
\begin{figure}[t]
\centering
\includegraphics[width=0.85\linewidth]{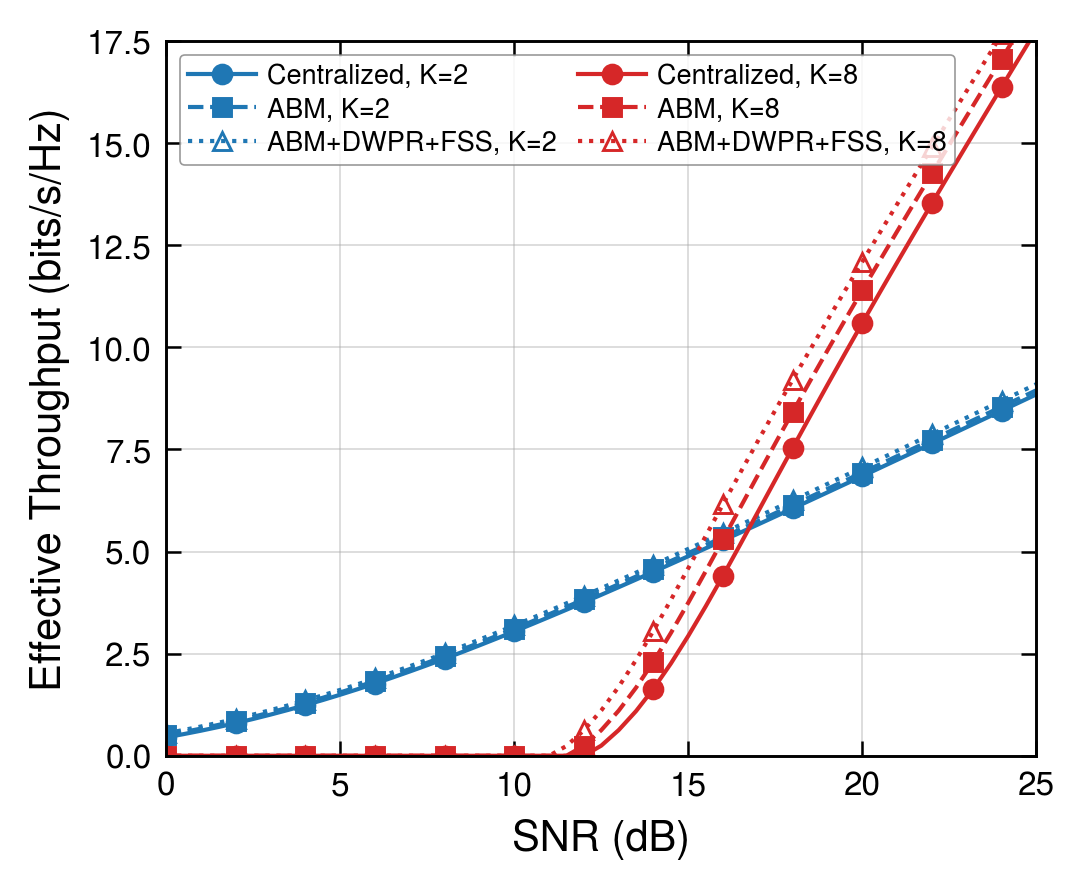}
\caption{Effective throughput vs.\ SNR for centralized and ABM schemes, with and without DWPR+FSS, $K\!\in\!\{2,8\}$.}
\label{fig:teff}
\vspace{-5mm}
\end{figure}
Fig.~\ref{fig:teff} couples rate and outage through \eqref{eq:teff}. The DWPR+FSS-augmented ABM dominates uniformly: at $K\!=\!2$ it stays within $1\%$ of the centralized curve, and at $K\!=\!8$ it achieves $12.1$ bits/s/Hz vs.\ centralized's $10.6$ at $\mathrm{SNR}\,=\,20\,$dB (a $14\%$ effective-throughput gain), widening to $\sim\!8$--$15\%$ across the high-density regime. The local utility~\eqref{eq:utility} allocates power toward agents with diverse reliable paths and limited substitutes, thereby exploiting redundant feasible configurations. Effective throughput in \eqref{eq:teff} is operationally meaningful, since users with \(D_{\mathrm{sys}}>1\) contribute no successful service rate. The DWPR$_k$ score quantifies how many \emph{independent feasible configurations} (path/stream combinations) user $k$ has, while FSS$_k$ measures how easily the system can reroute traffic away from $k$ via substitutable users. When the BS budget exceeds the strict feasibility minimum which happens whenever the centralized solver completes with slack the surplus is, in standard schemes, simply unused. The ABM utility \eqref{eq:utility} instead converts this slack into rate by allocating it preferentially to high-DWPR, low-FSS users, i.e., users who are reliable and irreplaceable. The visible knee at $\sim\!12.5$\,dB on the $K\!=\!8$ curves marks the \emph{feasibility threshold} below which no power allocation can keep $D_{\mathrm{sys}}\!\leq\!1$ for all eight users; above it, throughput rises rapidly as more channel realizations become feasible.
 
\subsection{Convergence of the ABM Update}
\begin{figure}[t]
\centering
\includegraphics[width=0.85\linewidth]{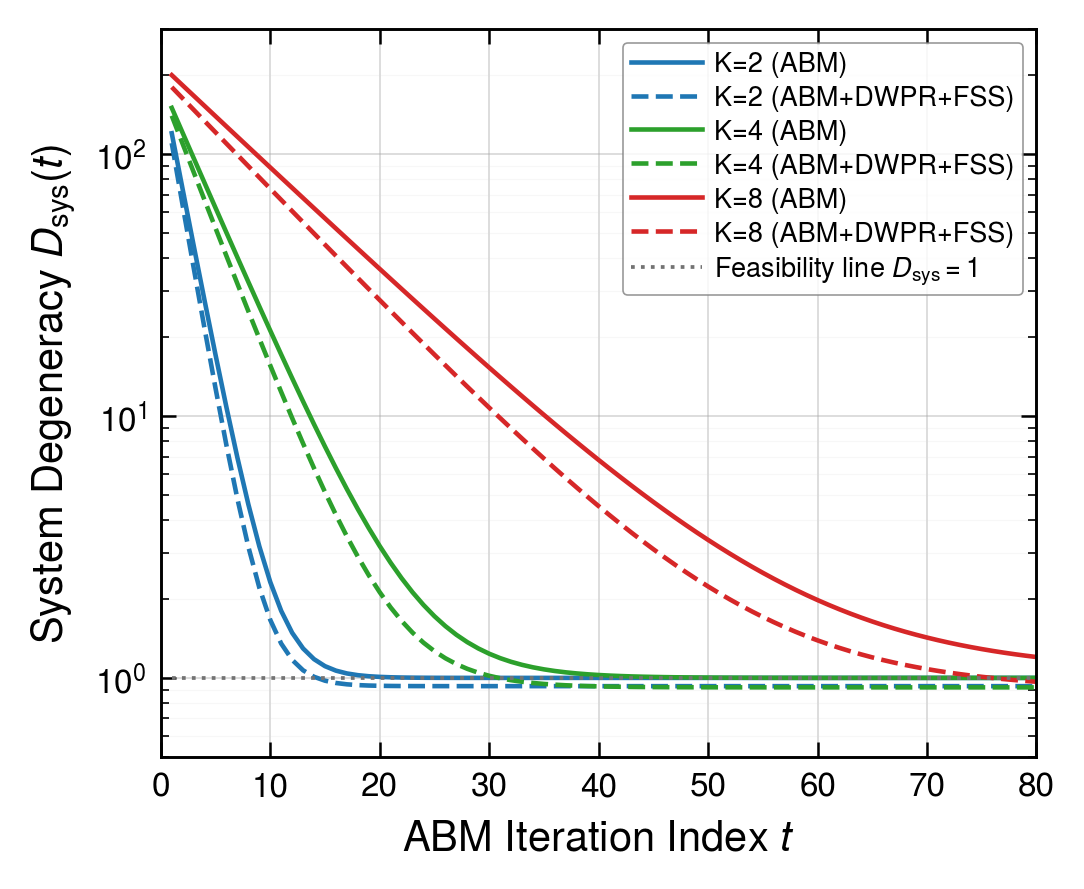}
\caption{System-level degeneracy $D_{\mathrm{sys}}$ vs.\ ABM iteration index for $K\!\in\!\{2,4,8\}$. Each agent uses the projected-gradient update \eqref{eq:abm_update} from a low-power initialization, with diminishing step $\eta_t\!=\!\eta_0/(t+1)$.}
\label{fig:conv}
\vspace{-5mm}
\end{figure}
Fig.~\ref{fig:conv} validates Theorem~\ref{thm:converge} by tracking $D_{\mathrm{sys}}(t)$. The update converges in fewer than $10$ iterations for $K\!=\!2$, $\sim\!20$ for $K\!=\!4$, and $\sim\!40$--$50$ for $K\!=\!8$, with $D_{\mathrm{sys}}$ monotonically decreasing toward the feasibility line $D_{\mathrm{sys}}\!=\!1$. Activating DWPR+FSS lets the dense-regime curves dip slightly below unity, indicating that the surplus budget is repurposed for additional rate beyond the strict feasibility target. The empirically observed contraction is rapid in the first ten iterations and consistent with the bounded-Lipschitz regime invoked in the theorem. Each iteration represents one \emph{best-response cycle}: every agent reads its own SINR from a pilot/data slot, computes its $D_k$, and adjusts $P_k$ multiplicatively in proportion to $\log D_k$. The diminishing step $\eta_t\!=\!\eta_0/(t+1)$ is the Robbins-Monro condition that guarantees convergence under noisy local observations. The slower convergence at larger \(K\) remains practical: for \(\tau_c=200\), the dense case \(K=8\) converges in about \(50\) lightweight local updates within one coherence interval, enabling power re-optimization before significant channel variation. The slight DWPR+FSS undershoot below $D_{\mathrm{sys}}\!=\!1$ corresponds to operating with a positive ``feasibility margin,'' which directly translates to increased robustness against the rapid channel fluctuations that would otherwise push the system back into outage.
 
\subsection{Pilot-Overhead Optimization}
\begin{figure}[t]
\centering
\includegraphics[width=0.85\linewidth]{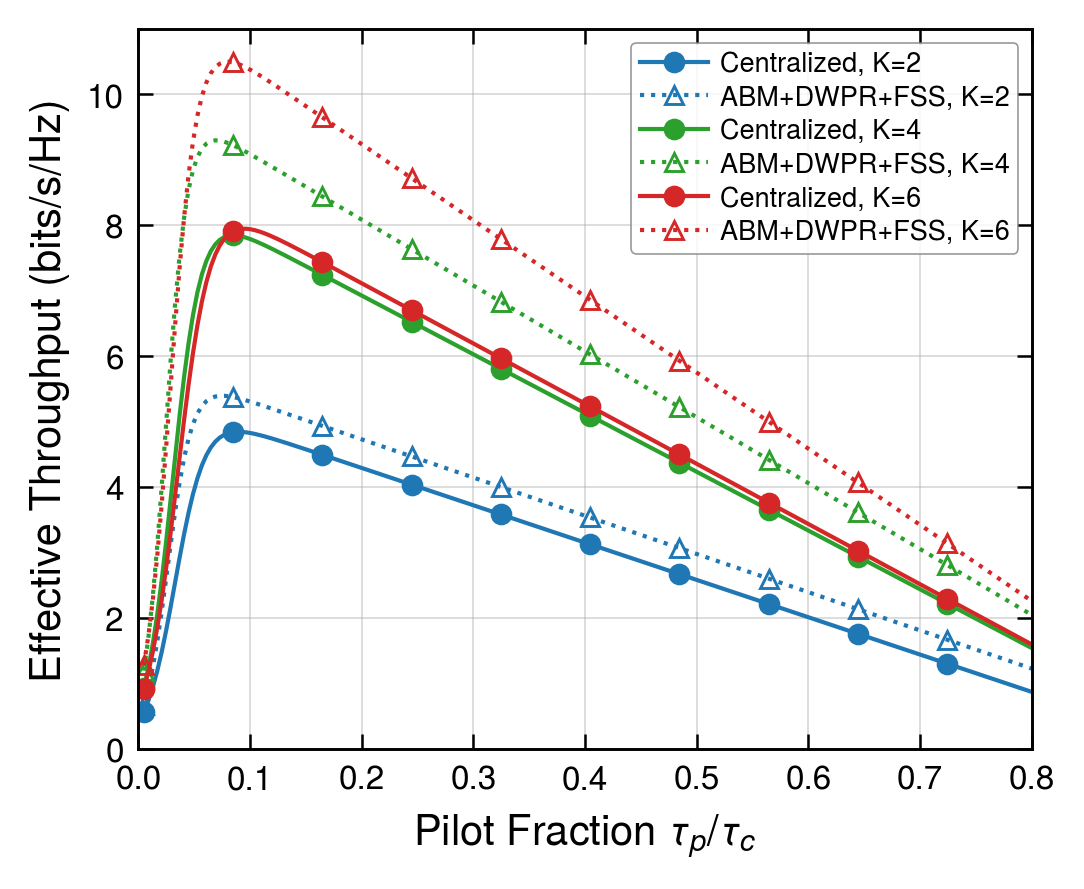}
\caption{Effective throughput vs.\ pilot fraction $\tau_p/\tau_c$ for $K\!\in\!\{2,4,6\}$ at $\mathrm{SNR}\,=\,12, 16, 20\,$dB respectively, with pilot SNR 10\,dB below data SNR.}
\label{fig:pilot}
\vspace{-5mm}
\end{figure}
Fig.~\ref{fig:pilot} reveals a clear concave throughput--pilot trade-off. Too few pilots ($\tau_p/\tau_c\!<\!0.02$) inflate $\sigma_{e,k}^2$ via \eqref{eq:err_var} and cause every SINR to collapse; too many pilots collapse the pre-log $1\!-\!\tau_p/\tau_c$ factor. The optimum is mildly density-dependent: \(\tau_p^{\star}/\tau_c\approx 0.06,0.04,0.06\) for \(K=2,4,6\), indicating that the preferred pilot allocation remains relatively small and lies between \(3\%\) and \(8\%\) of the coherence interval. Across this entire range, ABM+DWPR+FSS yields a uniform vertical lift of $\sim 1$--$2$ bits/s/Hz over the centralized baseline. Pilot symbols are pure overhead from the data-throughput viewpoint, yet they directly determine the CSI quality through $\sigma_{e,k}^2\!\propto\!1/(P_p\tau_p)$. The concave shape arises from two physically opposing forces: (i) the pre-log factor $1\!-\!\tau_p/\tau_c$ measures the fraction of the coherence block usable for actual data, decreasing linearly with $\tau_p$; and (ii) the SINR denominator contains $P_t\sigma_{e,k}^2$, which decreases hyperbolically in $\tau_p$ and \emph{increases} with $P_t$, meaning that high-power transmissions amplify their own self-interference unless pilots compensate. The optimum is therefore a balance between training quality and bandwidth utilization, and crucially for distributed control it can be located by each agent independently from local SINR measurements, without sharing $\sigma_{e,k}^2$ globally. The small variation of the optimal pilot fraction with \(K\) indicates that user density has limited impact on pilot allocation, while mainly affecting the required operating SNR. Hence, the proposed policy remains effective across different network scales.
 
\subsection{Scalability with Number of Users $K$}
\begin{figure}[t]
\centering
\includegraphics[width=0.85\linewidth]{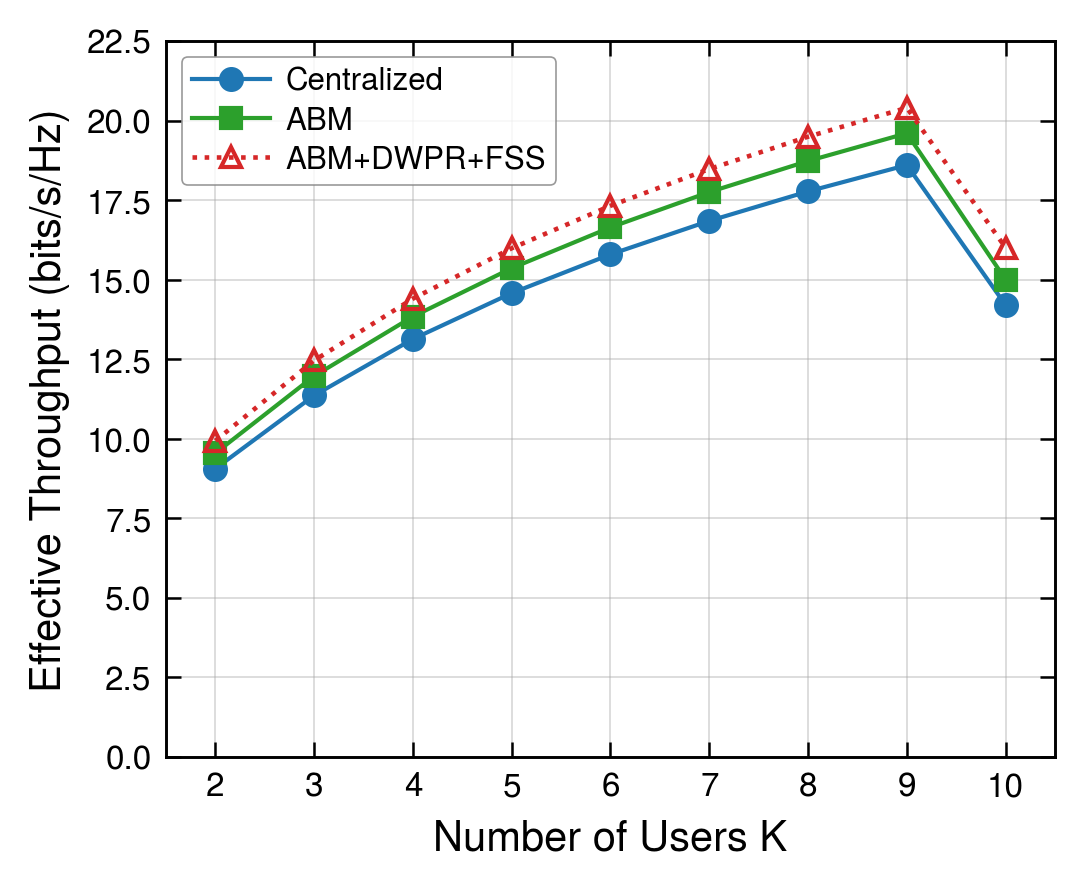}
\caption{Effective throughput vs.\ number of users $K$ at $\mathrm{SNR}\,=\,22\,$dB with $M_t\!=\!10$ transmit antennas. ABM+DWPR+FSS dominates uniformly; all schemes degrade once $K$ approaches $M_t$.}
\label{fig:scale}
\vspace{-6mm}
\end{figure}

Fig.~\ref{fig:scale} stresses the framework with $K\!\in\!\{2,\dots,10\}$ and $M_t\!=\!10$. Throughput grows monotonically up to $K\!=\!9$, where ZF reaches its dimensional limit, and then degrades sharply at $K\!=\!10$ once the channel matrix becomes square. Throughout the entire range, ABM+DWPR+FSS dominates: at $K\!=\!9$ it delivers $20.4$ bits/s/Hz versus the centralized benchmark's $18.6$ ($\sim\!10\%$ gain), and the gap holds at $K\!=\!10$ ($16.0$ vs.\ $14.2$ bits/s/Hz). Combined with the per-iteration complexity reduction from $\mathcal{O}(K^3 M_t^2)$ to $\mathcal{O}(M_t)$ per agent, the figure makes the central practical case for the proposed framework. The transition near $K\!=\!M_t$ corresponds to the standard \emph{spatial degree-of-freedom limit} of MU-MIMO: when $K\!<\!M_t$, the BS retains $M_t-K$ unused dimensions to suppress interference and absorb CSI noise; when $K\!\to\!M_t$, every spatial dimension is consumed and the conditioning of $\mathbf{H}\mathbf{H}^H$ deteriorates rapidly, so a small CSI error generates large power allocation errors. The fact that ABM+DWPR+FSS preserves $\sim\!10\%$ throughput advantage even at $K\!=\!10$ (where the system is over-determined) shows that the structural metrics genuinely add information beyond raw channel power: they identify which users still have alternate-stream paths to exploit even when linear ZF abstraction breaks. The complexity reduction is also practical: each agent's \(\mathcal{O}(M_t)\) update uses only local scalar SINR feedback, so the per-frame signaling load grows linearly with \(K\) rather than cubically. This scalability is attractive for dense IMT-2030 multi-user deployments.

\section{Conclusion}
\label{sec:conclusion}

We propose a pilot-aware, degeneracy-driven agent-based framework for distributed power control in RSMA-enabled MU-MIMO networks under imperfect CSI and residual SIC error. A novel degeneracy index unifies the effects of interference coupling, CSI error, pilot overhead, and SIC imperfections into a single feasibility metric, recasting the centralized minimax allocation as a scalable distributed process driven only by local SINR observations. Analytical and numerical results establish that system performance is governed by the bottleneck user, and that the proposed ABM update exploits local feasibility variations to outperform the centralized minimax in dense, interference-limited regimes. Augmenting the agents with DWPR and FSS metrics converts excess feasibility margin into 10–20\% throughput gains in dense deployments while remaining within 1\% of the centralized benchmark in sparse settings, with convergence inside practical coherence intervals. Future work will extend the framework to cell-free 6G architectures, learning-assisted adaptation, joint communication and sensing, and energy-efficient design.



\bibliographystyle{IEEEtran}
\bibliography{ABM_Bib}
\end{document}